\documentclass[prl,aps,epsf,twocolumn,superscriptaddress,longbibliography]{revtex4-1}

\usepackage{amsmath,amssymb,amsfonts,bbm,float,graphics,epsfig,epstopdf,color,verbatim,tabularx,bm,multirow,hyperref,ulem,times,subfigure}
\hypersetup{
  colorlinks, linkcolor={blue},
  citecolor={blue}, urlcolor={blue}
}

\usepackage{textcomp}
\usepackage{times}
\usepackage{tabularx}
\usepackage{graphicx}
\usepackage{float}
\usepackage{latexsym,amsmath,amssymb,bm,euscript}
\usepackage{color}
\usepackage{subfigure}
\usepackage{epstopdf}

\graphicspath{{./Figs/}}

\begin{document}

\title{Continuous Transition between Bosonic Fractional Chern Insulator and Superfluid}
\author{Hongyu Lu}
\affiliation{Department of Physics and HK Institute of Quantum Science \& Technology, The University of Hong Kong, Pokfulam Road, Hong Kong, China}
\author{Han-Qing Wu}
\affiliation{Guangdong Provincial Key Laboratory of Magnetoelectric Physics and Devices, School of Physics, Sun Yat-sen University, Guangzhou 510275, China }
\author{Bin-Bin Chen}
\email{bchenhku@hku.hk}
\affiliation{Department of Physics and HK Institute of Quantum Science \& Technology, The University of Hong Kong, Pokfulam Road, Hong Kong, China}
\author{Zi Yang Meng}
\email{zymeng@hku.hk}
\affiliation{Department of Physics and HK Institute of Quantum Science \& Technology, The University of Hong Kong, Pokfulam Road, Hong Kong, China}

\begin{abstract}
The properties of fractional Chern insulator (FCI) phases and the phase transitions between FCI and Mott insulators (MI)  in bosonic systems are well studied. 
The continuous transitions between FCI and superfluid (SF), however, despite the inspiring field theoretical predictions~\cite{Barkeshli2012CFL_FL, Barkeshli2014FQH-SF,Barkeshli2015FCI-SF, Song2023deconfined, songPhase2024}, have not been directly verified. The existing numerical results of the FCI-SF transition are either indirect or clearly first-order. Here, by simply tuning the bandwidth of the Haldane honeycomb lattice model, we find direct transitions from a bosonic FCI at $\nu=1/2$ filling of a flat Chern band to two SF states with bosons condensed at momenta $M$ or $\Gamma$, respectively. 
While the FCI-SF($M$) transition is first-order, the FCI-SF($\Gamma$) transition is found continuous, and the bipartite entanglement entropy at the critical point with the area-law scaling is consistent with the critical theories. 
Through finite size criticality analysis, the obtained critical exponents $\beta\approx0.35(5)$ and $\nu\approx0.62(12)$ are both compatible with those of the 3D XY universality class within numerical uncertainty and possibly more exotic beyond-Landau ones.
This letter thence presents a {\it direct} numerical demonstration of a continuous FCI-SF transition 
between topologically ordered phase and spontaneous continuous symmetry-breaking phase,
and further indicates the zero-field bosonic FCI might be realized from a SF state by gradually flattening the dispersion of the Chern band, through the (quasi)adiabatic preparation in ultracold atom systems. 
\end{abstract}

\date{\today }
\maketitle

\noindent{\textcolor{blue}{\it Introduction.}---}
The fractional quantum Hall (FQH) effect is one of the focal topics of strongly correlated systems in the past decades~\cite{Stormer1999FQH, Klaus2020FQH, Laughlin1983FQH, Haldane1983FQH, Jain1992FQH, GMP1985,jainComposite2007,Haldane2011FQHgeometrical, kukushkinDispersion2009, Lin2014FQHex}, whose low-energy physics is governed by the Chern-Simon gauge
theory~\cite{Fradkin1989CS, Kivelson1989CS, Fradkin1991chernsimons, Wen1995CS, Sun2015CS}. 
Further, fractional Chern insulators (FCIs) without magnetic field have been proposed~\cite{Sun2011_fci, DNSheng2011_fci, DNSheng2011boson, Tang2011FCI, Titus2011_fci, Bernevig2011_fci, LFu2021_fci, Heqiu2021_fci, Neupert_2015FCI} and recently realized in experiments of Moir\'e materials~\cite{caiSignature2023,park2023_fqah,zengThermodynamic2023,xu2023_fci, kang2024fqsh,multilayer_graphene_fqah}. 

In bosonic systems, although the ground-state properties of FQH/FCI states are well studied~\cite{Cooper2020FQH, DNSheng2011_fci, Zhu2016FQAH, Rosson2019FQH, Luo2020BosonFCI, Zeng2022FQH}, the knowledge of related phase transitions are relatively limited, especially in concrete lattice model systems. 
Despite the well-understood continuous transition between Mott insulator (MI) and superfluid (SF)~\cite{Fisher1989localization, Greiner2002SF-Mott} and the existing studies of continuous transitions between FQH/FCI states and
MI~\cite{Kivelson1989CS,Wen1993FQH-Mott, Chen1993Mott, Kivelson1992Mott, Barkeshli2014FQH-SF,Motruk2017transition, zeng2021transition}, an important question still remaining open is whether the transition between FQH/FCI (topological order) and SF (Landau order) can be continuous~\cite{Motruk2017transition, Valentin2020FCI_SF, zeng2021transition}. Theoretically, this could be another example of transitions beyond the Landau-Ginzburg paradigm~\cite{Barkeshli2012CFL_FL, Barkeshli2014FQH-SF, Barkeshli2015FCI-SF,songPhase2024,senthil2004DQCP, senthilQuantum2004,liuDisorder2024,slagleExotic2015,ayyarMassive2015, Wang2017DQCP,qinDuality2017} which could require emergent gauge field and fractional excitations, and render exotic consequences such as continuum in spectra~\cite{maDynamical2018}, universal jump in resistivities~\cite{senthilTheory2008,songPhase2024} and fractionalization of symmetries~\cite{sunDynamical2018}, compared to more well-understood continuous quantum phase transitions between symmetric and symmetry-breaking phases~\cite{Sondhi1997continuous, Sachdev1999transition}.  

Although the field theoretic proposals of such continuous transitions have been put forward~\cite{Barkeshli2012CFL_FL,Barkeshli2014FQH-SF, Barkeshli2015FCI-SF, Song2023deconfined, songPhase2024}, the proposed supporting evidence of possibly continuous FCI-SF transition by tuning the band dispersion of a square-lattice model is only indirect~\cite{Barkeshli2015FCI-SF}, where the authors perturbatively break the suggested quantum-critical-point-protecting inversion symmetry 
and introduce an intermediate MI state. Therefore, the direct proof of possible continuous FCI-SF transition is still lacking. Other attempts of finding the direct FCI-SF transitions, such as by softening the Hubbard repulsion in a topological flat band model~\cite{zeng2021transition}, tuning the band topology through staggered potential~\cite{Motruk2017transition}, and changing the interaction strength in a coupled-wire model~\cite{Valentin2020FCI_SF},
 appear to be first-order.

Experimentally, ultracold systems on optical lattices are promising avenues to realize bosonic FQH/FCI states~\cite{anders2005opticalFQH, Palmer2006opticalFQH,Palmer2008opticalQH,Yao2013dipolar, Cooper2013optical}. However, in such quantum optical systems, it could be more difficult to reach the ground state than to realize the appropriate Hamiltonian~\cite{Miyake2013Harper, Aidelsburger2013Hofstadter, Aidelsburger2015hofstadter}. 
One potential challenge is that it might be difficult to cool down the system by an external bath, and to circumvent it, the (quasi)adiabatic preparation scheme is invented and believed to overcome the problem~\cite{Popp2004adiabatic, Schaller2006adiabatic, Barkeshli2015FCI-SF,He2017adiabatic, Motruk2017transition, zeng2021transition, Carcy2021adiabatic}. This is because in bosonic lattices, SF and MI states are simpler to realize experimentally and could serve as an initial state for adiabatic evolution. Indeed, FQH states have been realized in finite systems with a few bosonic atoms or photons through adiabatic preparation from localized states~\cite{Julian2023photonFQH, Wang2024photonFQH}. 
Therefore, the direct evidence of continuous FCI-SF transition might provide further perspectives for the realization of zero-field bosonic FCI in even larger Chern-band systems, which also motivates us to verify such theoretical predictions via large-scale numerical simulations.

\begin{figure}[htp!]
\centering		
\includegraphics[width=0.5\textwidth]{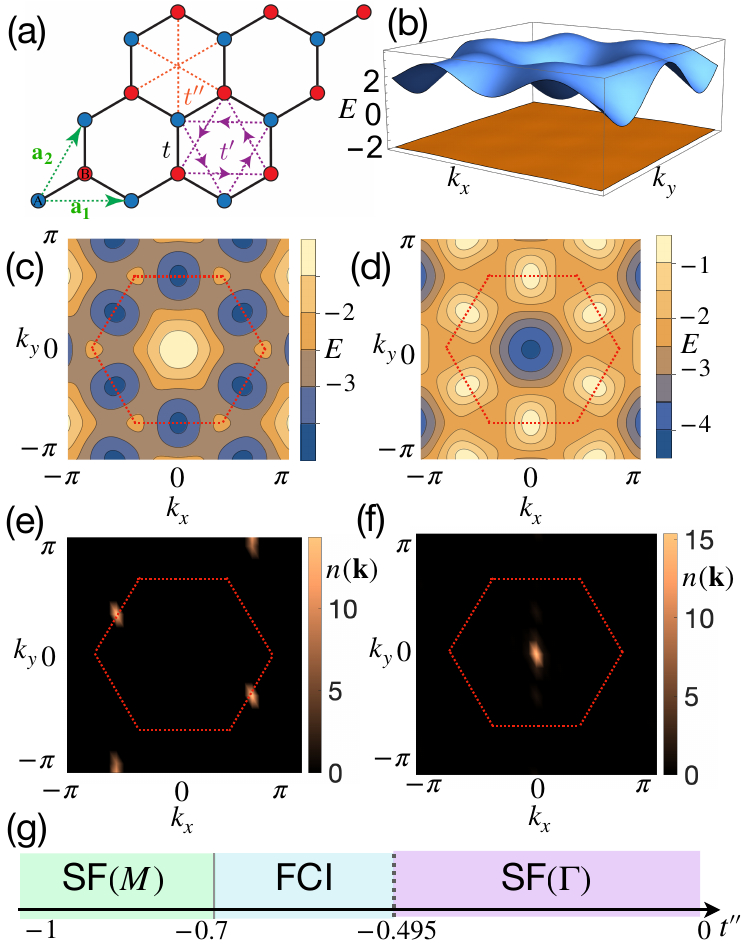}
\caption{ \textbf{Model and phase diagram.} (a) The honeycomb lattice with zigzag geometry ($N=3\times3\times2$ for example here), where blue(red) sites refer to A(B) sublattice, respectively. (b) The single-particle energy bands with the flat-band parameters in Eq.~\eqref{eq:eq1}.  We show the contour plot of the lower single-particle Chern band with (c) $t''=-1$ and (d) $t''=0$, which are dispersive. The energy minimum is at (c) $M$ point and (d) $\Gamma$ point, respectively. Boson occupation $n(\mathbf{k})$ in the Brillouin zone (BZ) with (e) $t''=-1$ and (f) $t''=0$ from DMRG simulations, and the Fourier transformation is done using the central $N_y\times N_y=10\times10$ sites in the bulk. In the two SF phases, the condensed momenta are in agreement with the single-particle energy minima. The red dotted lines in (c-f) represent the first BZ. (g) Phase diagram when the lower Chern band is half filled with hard-core bosons when tuning $t''$ with $V_1=V_2=0$. The gray solid line represents a first-order FCI-SF($M$) transition, while the gray dashed line represents the continuous FCI-SF($\Gamma$) transition.} 
	\label{fig_fig1}
\end{figure}

In this work, we use density matrix renormalization group (DMRG)~\cite{While1992dmrg, dmrg2005review} and exact diagonalization (ED)~\cite{2008ed} methods to study the Haldane model on the honeycomb lattice~\cite{HaldaneModel1988} with hard-core bosons half filling the lower Chern band. 
When the Chern band is flat, the $\nu=1/2$ FCI ground state can be robustly realized even without further neighboring interactions~\cite{DNSheng2011boson, Luo2020BosonFCI}. 
We start from the flat-band setting and gradually tune the band dispersion with a single hopping parameter. 
We find two quantum phase transitions: a first-order transition from FCI to a SF state with bosons condensed at one $M$ point (broken $C_3$ rotational symmetry), dubbed SF($M$); and a continuous transition from FCI to a SF state with bosons condensed at $\Gamma$ point, dubbed SF($\Gamma$). 
We further analyze the quantum critical behavior of the FCI-SF($\Gamma$) transition through approximately (2+1)D finite-size scaling analysis and find the critical exponents $\beta\approx0.35(5)$ and $\nu\approx0.62(12)$. Within numerical uncertainty, they might be compatible with those of the 3D XY universality class while remaining possible to be more exotic beyond-Landau ones. We note, to completely determine whether the exact values of critical exponents should deviate from those of the 3D XY universality class, more accurate simulations are still needed.
Besides, the behavior of bipartite entanglement entropy at the quantum critical point satisfies the area law, consistent with that of (2+1)D critcal theories~\cite{Barkeshli2014FQH-SF,songPhase2024,fradkinEntanglement2006,songExtracting2024,liuDisorder2024}.
Moreover, we find that the FCI-SF($\Gamma$) transition through tuning band dispersion is also robustly continuous in the presence of different neighboring repulsive interactions (even when they are much larger than the single-particle band gap). 
To the best of our knowledge, these results provide the long-sought {\it direct} evidence and criticality analysis for continuous FCI-SF transition, and we propose such a transition could provide more perspectives for the experimental realization of zero-field bosonic FCI in ultracold systems through (quasi)adiabatic preparation.

\noindent{\textcolor{blue}{\it Model and Methods.}---} We study the Haldane honeycomb-lattice model~\cite{HaldaneModel1988} with extended hoppings:
\begin{equation}
	\begin{aligned}
		H =&-\sum_{\langle i,j\rangle}t(b_i^\dagger b^{\,}_j+\text{H.c.})-\sum_{\langle\hskip-.5mm\langle i,j \rangle\hskip-.5mm\rangle}t'(e^{i\phi}b_i^\dagger b^{\ }_j+\text{H.c.})\\
		&-\sum_{\langle\hskip-.5mm\langle\hskip-.5mm\langle i,j \rangle\hskip-.5mm\rangle\hskip-.5mm\rangle} t''(b_i^\dagger b^{\ }_j+\text{H.c.})+V_1\sum_{\langle i,j\rangle}n_in_j+V_2\sum_{\langle\hskip-.5mm\langle i,j \rangle\hskip-.5mm\rangle}n_in_j,
	\end{aligned}
	\label{eq:eq1}
\end{equation}
where $b_i^\dagger$ ($b_i$) creates (annihilates) a hard-core boson at the $i$-th site. We consider a zigzag geometry as shown in Fig.~\ref{fig_fig1} (a) and set nearest-neighbor (NN) $t=1$, next-nearest-neighbor (NNN) $t'=0.6$, next-next-nearest-neighbor (NNNN) $t''=-0.58$ and $\phi=0.4\pi$, which are found as optimal flat-band parameters in this model~\cite{DNSheng2011boson} and the Chern bands at this setting are shown in Fig.~\ref{fig_fig1} (b). In this work, we focus on fixed $t$ and $t'$ while tuning only $t''$ to control the bandwidth of the lower band. $V_1$ ($V_2$) refers to the amplitude of NN (NNN) repulsive interactions. We define the primitive vectors $\mathbf{a_1}=(\sqrt{3},0)$ and $\mathbf{a_2}=(\frac{\sqrt{3}}{2},\frac{3}{2})$.

The results in the main text are from DMRG simulations with U(1) symmetry of the conserved total number of bosons, and we consider finite-length cylinders with $N_y\times N_x$ unit cells and total lattice sites $N=N_y\times N_x\times2$. We focus on the half filling of the lowest Chern band $\nu=N_\mathrm{b}/(N_y\times N_x)=1/2$ where $N_\mathrm{b}$ is the number of bosons. 
We consider $N_y=4, 6, 8, 10$ respectively for the periodic direction along $\mathbf{a}_2$. For the length of cylinders, we consider up to $N_x=30$. 
We keep the bond dimensions up to $D=4096$, and the maximum truncation error for $N_y=10$ is of order $10^{-5}$. 

\noindent{\textcolor{blue}{\it Phase diagram with $V_1=V_2=0$.}---}
We mainly focus on the hard-core bosons without neighboring interactions ($V_1=V_2=0$), where the robust FCI ground state at the half-filled flat Chern band ($t''=-0.58$) is numerically verified~\cite{DNSheng2011boson,Luo2020BosonFCI}. 
When increasing the magnitude of NNNN hopping from $|t''|=0.58$, the single-particle energy minimum of the dispersive lower band locates at the $M$ point of the hexagonal BZ, while it is at the $\Gamma$ point when decreasing $|t''|$ from the flat-band setting, as shown in Fig.~\ref{fig_fig1} (c) and (d), respectively. The ground-state phase diagram obtained from DMRG simulation is shown in Fig.~\ref{fig_fig1} (g). 
When the Chern band gradually goes dispersive, the resulting state from FCI is either SF($M$) or SF($\Gamma$), where the condensed momenta are in agreement with the single-particle dispersion. We plot the boson occupation $n(\mathbf{k})=n_\alpha(\mathbf{k})=\frac{1}{N}\sum_{i,j}e^{-i\mathbf{k}\cdot\mathbf{r_{i,j}}}\langle b_{i,\alpha}^\dagger b^{\ }_{j,\alpha}\rangle$ (where $\alpha=A/B$ specifies the sublattices, and we sum over only B sublattice for symmetry simplicity in the bulk region that consists of $N_y\times N_y$ sites )
in the BZ in Fig.~\ref{fig_fig1} (e) and  (f) at $t''=-1$ and $t''=0$ respectively to illustrate the SF phases. In our DMRG results of the SF($M$) state, the bosons spontaneously condense at a single $M$ point, and the $C_3$ rotational symmetry is broken.

\begin{figure}[htp!]
	\centering		
	\includegraphics[width=0.5\textwidth]{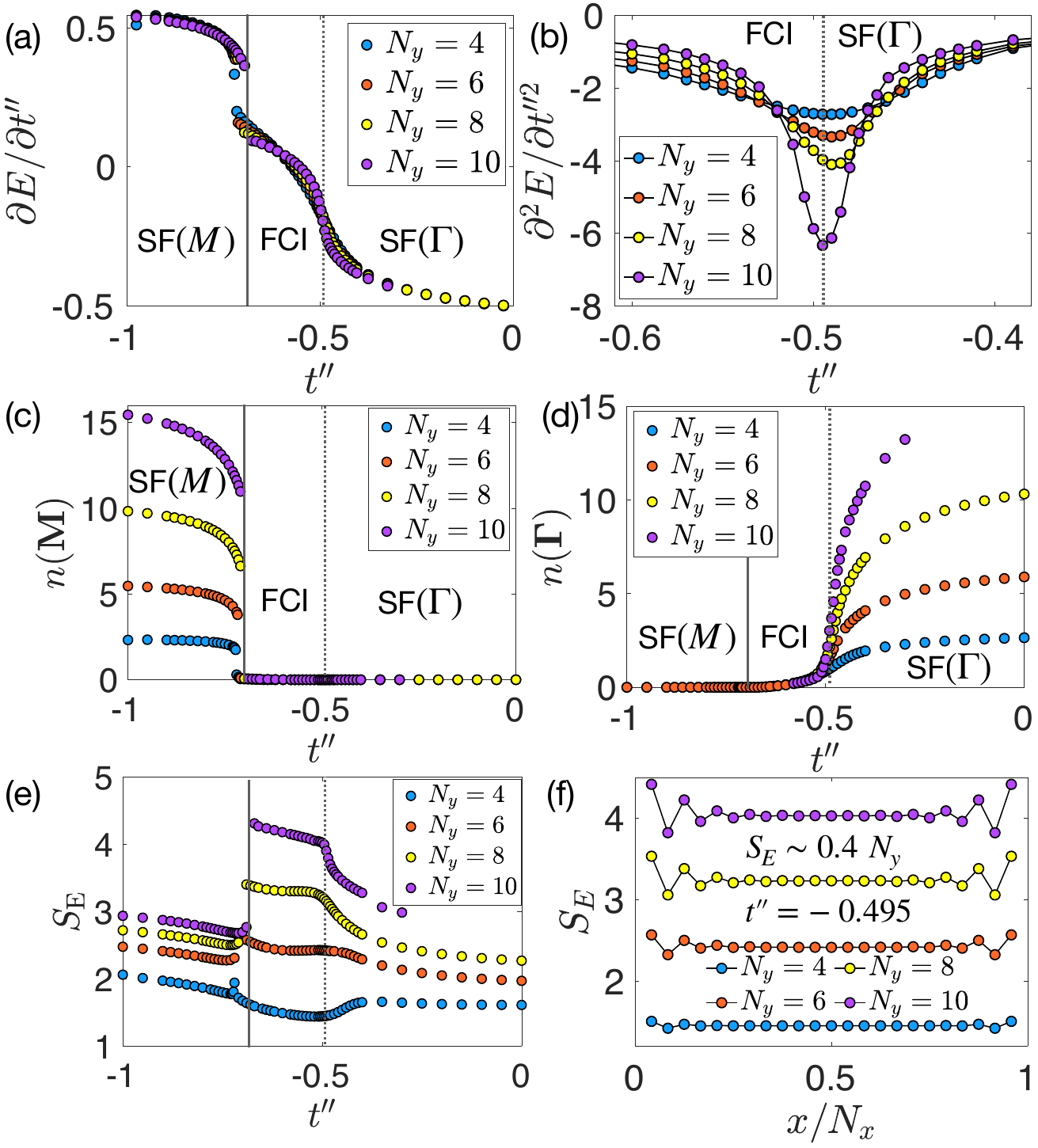}
	\caption{ \textbf{Nature of the quantum phase transitions.} The DMRG results of (a) first derivative of per-site energy $\frac{\partial E}{\partial t''}$, (b) second derivative of per-site energy $\frac{\partial^2 E}{\partial t''^2}$, boson occupations at the (c) $M$ point and (d) $\Gamma$ point, and (e) bipartite entanglement entropy $S_E$ as a function of $t''$ from cylinders up to $N_y=10$, are shown. We use the grey solid and dashed lines to label the first-order FCI-SF($M$) transition at $t''\approx -0.7$ and the continuous FCI-SF($\Gamma$) transition at $t''_c \approx -0.495$, respectively in these panels. (f) The scaling of $S_E$ at critical point $t''\approx0.495$ for different $N_y$, which is agreement with the area law without the logarithm dependence of $x$.}
	\label{fig_fig2}
\end{figure}

To  elucidate the nature of the quantum phase transitions in Fig.~\ref{fig_fig1} (g), we show the detailed DMRG results in Fig.~\ref{fig_fig2}. 
We plot the first and second derivatives of per-site energy $\frac{\partial E}{\partial t''}$ and $\frac{\partial^2 E}{\partial t''^2}$, the boson occpation $n(\mathbf{M})$ and $n(\mathbf{\Gamma})$, and bipartite entanglement entropy $S_E$ as functions of $t''$ from DMRG simulations in Fig.~\ref{fig_fig2} (a-d). 
The discontinuity of the FCI-SF($M$) transition is getting more clear when increasing the width of cylinders, as shown from $\frac{\partial E}{\partial t''}$, $n(\mathbf{M})$, and $S_E$ (the first-order nature is rather weak for the small $N_y=4$). While the discontinuous transition point is slightly drifting, we directly take the value ($t''\approx-0.7$) from the data of $N_y=10$ for this first-order transition.
On the other side, the continuous behavior of $\frac{\partial E}{\partial t''}$ and $S_E$ as well as the condensed bosons $n(\mathbf{\Gamma})$ for all considered system sizes support the continuous nature of the FCI-SF($\Gamma$) transition.
Besides, for this continuous transition, the peak of $\frac{\partial^2 E}{\partial t''^2}$ becomes sharper and more pronounced as the system size increases [Fig.\ref{fig_fig2}(b)].
Moreover, we show the scaling of entanglement entropy at the critical point $t''=-0.495$ in Fig.~\ref{fig_fig2}(f). Here, our DMRG simulations consider cylinders with fixed $N_x=24$ and increasing $N_y$. Interestingly, at critical point, there is no quasi-1D logarithm dependence of $x$ for all system sizes, and the entanglement entropy scales with area law with smooth boundary~\cite{Srednicki1993area, Eisert2010critical,fradkinEntanglement2006,songExtracting2024,liuDisorder2024}, which is consistent with existing (2+1)D critical theories with Dirac cones~\cite{Barkeshli2014FQH-SF,songPhase2024}.
We provide supplementary information in the SI~\cite{suppl}.

\begin{figure}[htp!]
	\centering		
	\includegraphics[width=0.5\textwidth]{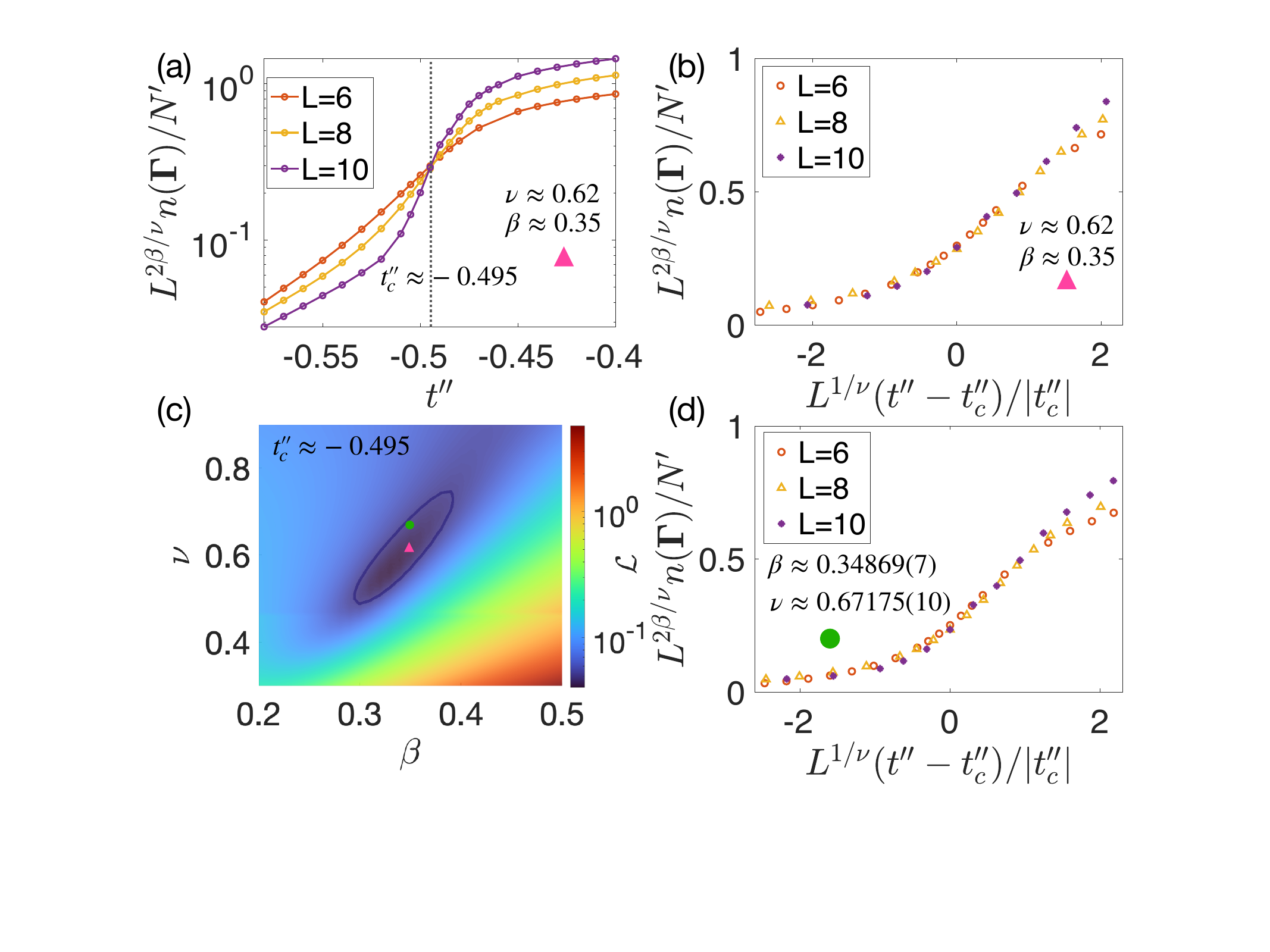}
	\caption{ \textbf{Finite-size criticality analysis of FCI-SF($\Gamma$) transition.} (a) Rescaled per-site occupation at $\Gamma$ point of the BZ $n(\mathbf{\Gamma})/N'$ as a function of $t''$, which show good crossing at $t''_\mathrm{c}\approx-0.495$. (b) Scaling collapse obtained by plotting $L^{2\beta/\nu}n(\mathbf{\Gamma})/N'$ as a function of $L^{1/\nu}(t''-t''_c)/|t''_c|$. 
	The critical exponents in (a) and (b) are from (c), which shows the loss function of data collapse with changing $\beta$ and $\nu$, defined as the squared deviation of the fitted scaling function away from the data points. 
	With $t''_c\approx0.495$, the optimal critical exponents exist in a range of parameter space with $\beta\approx0.35(5)$ and $\nu\approx0.62(12)$, denoted by the deep blue contour, which is from the nearest extremum of the loss function gradient. 
	The pink triangle labels the exponents used in (a) and (b), while the green circle labels the critical exponents of the 3D XY universality class, directly using which we further show the data collapse in (d).
}
	\label{fig_fig3}
\end{figure}

\noindent{\textcolor{blue}{\it Quantum criticality of the FCI-SF($\Gamma$) transition.}---} To further characterize the continuous FCI-SF($\Gamma$) transition, we proceed with the finite size scaling analysis. 

To approximate the (2+1)D finite size scaling analysis, when obtaining $n(\mathbf{\Gamma})$ by taking the Fourier transformation of $\langle b_{i,B}^\dagger b^{\ }_{j,B}\rangle$, we choose a 2D bulk region including $N'=L\times L$ lattice sites where $L=N_y$ depending on the width of the cylinders. Therefore, we are scaling from $6\times6$ to $10\times10$ in the DMRG data [we do not use $N_y=4$ due to the more severe finite-size effect indicated by the relatively different behavior of entanglement entropy in Fig.\ref{fig_fig2}(e)]. 
Using the critical exponents $\nu\approx0.62$ and $\beta=0.35$, we plot the rescaled occupation at $\Gamma$ point of the BZ as a function of $t''$ in Fig.~\ref{fig_fig3} (a), with obvious crossing at $t''_\mathrm{c}\approx-0.495$.
We also show the scaling collapse of three system sizes obtained by plotting $L^{2\beta/\nu}n(\mathbf{\Gamma})/N'$ as a function of $L^{1/\nu}(t''-t''_c)/|t''_c|$ in Fig.~\ref{fig_fig3} (b), with good quality. 
The exponents in Fig.\ref{fig_fig3}(a,b) are obtained by optimizing the loss function of the finite-size data collapse, which is the squared deviation of the fitted scaling function away from the data points (the closer to the critical point, the higher weight for the deviation)~\cite{chenPhases2023}.
As shown in Fig.~\ref{fig_fig3} (c), the optimal critical exponents exist in a range of parameter space: $\beta\approx0.35(5)$ and $\nu\approx0.62(12)$, within numetical uncertainties. 

The critical theories of such a continuous transition consist of Dirac fermions coupled to a Chern-Simons gauge field~\cite{Barkeshli2014FQH-SF,Barkeshli2015FCI-SF,Song2023deconfined,songPhase2024}, but the knowledge of critical exponents is so limited.
If there are no gauge fluctuations as for a simple XY transition, the critical exponents are well known, including $\beta\approx0.34869(7)$ and $\nu\approx0.67175(10)$~\cite{Campostrini2001XY, Chester2020O2}. 
As shown in the loss function of critical exponents in Fig.\ref{fig_fig3}(c) and the data collapse using the critical exponents of the 3D XY universality class in Fig.\ref{fig_fig3}(d), the obtained optimal critical exponents are possibly consistent with those of the 3D XY universality class, since the difference is not so huge. But it remains possible that the exponents, especially $\nu$ might deviate from that of the 3D XY universality class and might need more accurate simulations.

\begin{figure}[htp!]
	\centering		
	\includegraphics[width=0.5\textwidth]{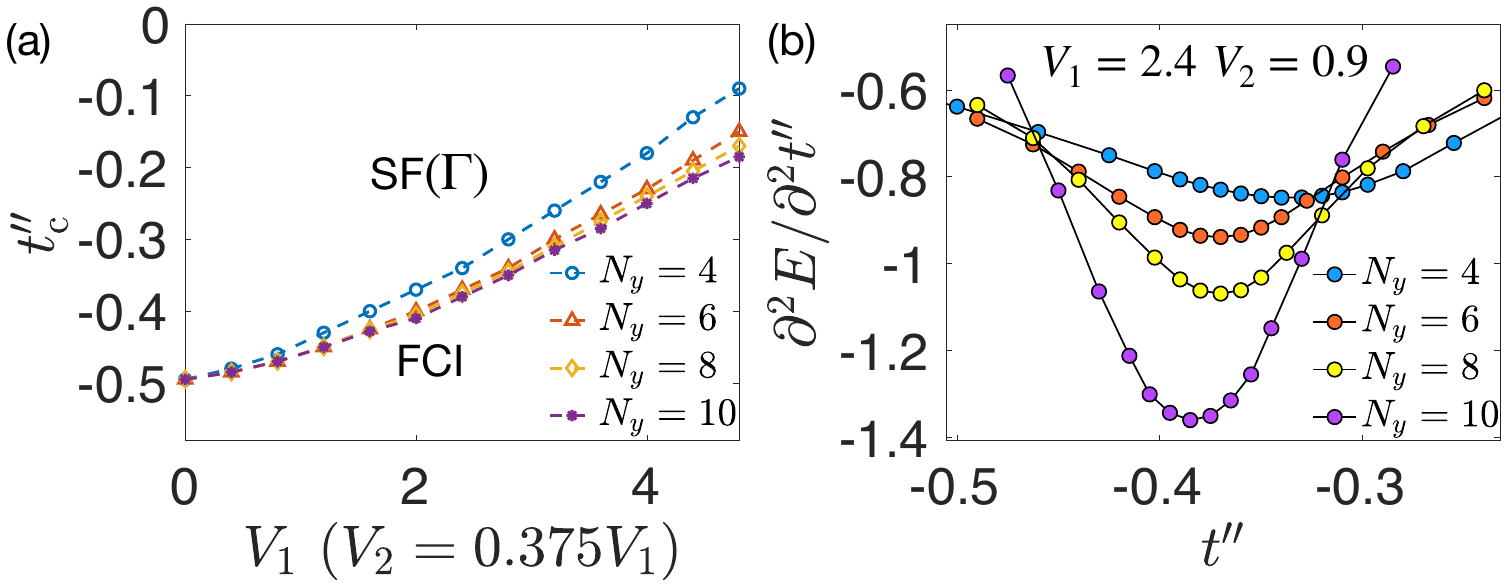}
	\caption{ \textbf{Continuous FCI-SF($\mathbf{\Gamma}$) transition with neighboring interactions.} 
		(a) The $V_1 (V_2=0.375V_1)-t''$ phase diagram shows the continuous FCI-SF($\Gamma$) transition is robust against neighboring interactions and critical line extends to strong $V_1/V_2$ regions. The critical boundaries at finite $V_1/V_2$ are determined by the minimum of second derivative of energy for each system size, and the labels refer to the different simulated values of $V_1/V_2$. 
		(b) The second derivative of per-site energy $\frac{\partial^2 E}{\partial t''^2}$ at $V_1=2.4$ and $V_2=0.9$ from cylinders of $N_y$ up to $10$.
	}
	\label{fig_fig4}
\end{figure}

\noindent{\textcolor{blue}{\it FCI-SF($\Gamma$) transition with neighboring interactions.}---}
In the above sections, we mainly focus on the hard-core bosons without neighboring interactions. 
We then study how robust this continuous FCI-SF($\Gamma$) is against neighboring interactions by studying a two dimensional $V_1(V_2)-t''$ phase diagram with fixed $V_2=0.375V_1$. The maximum interactions considered here is $V_1=4.8$ and $V_2=1.8$ (12 sets of gradually increasing interactions apart from $V_1=V_2=0$), since further increasing the $V_2$ interaction would lead to quantum phase transitions out of FCI even at the flat-band setting ($t''=-0.58$), but the nature of the larger-$V_2$ states is still under debate~\cite{DNSheng2011boson,Luo2020BosonFCI,Lu2024vestigial}. Therefore, we choose interaction regions where the ground-state at flat-band limit is robustly FCI.
The $V_1(V_2)-t''$ phase diagram confirmed from DMRG results of cylinders with width up to $N_y=4,6,8,10$ is shown in Fig.\ref{fig_fig4}(a). 
Interestingly, we find the continuous FCI-SF($\Gamma$) transition is robust against neighboring interactions even much larger than the single-particle band gap, and the Chern band is getting more dispersive at critical $t''_c$ when the interactions increase.
We show the second derivative of per-site energy $\frac{\partial^2 E}{\partial t''^2}$ with $V_1=2.4$ and $V_2=0.9$ in Fig.\ref{fig_fig4}(b) as an example to further demonstrate this continuous transition, and the peaks are getting sharper and more prominent when increasing $N_y$.
We also note that the peaks of $\frac{\partial^2 E}{\partial t''^2}$ are getting less pronuonced than the case without neighboring interactions [Fig.\ref{fig_fig2} (b)], suggesting the continuous transition might get more smooth when increasing interactions.
This $V_1(V_2)-t''$ phase diagram indicates that the (quasi)adiabatic preparation of FCI state in cold atom experiments might be possible with neighboring interactions as well, and the FCI could be realized even further away from the flat-band limit.

Besides the SF($M$) and SF($\Gamma$) states in this model, we also find another SF state with bosons condensed at the $K$ point of the BZ by introducing NN attractive interactions with the flat-band parameters. The interaction-driven transition from FCI to finite-momentum SF($K$) is just first-order and we discuss it in the SI~\cite{suppl}.

\noindent{\textcolor{blue}{\it Discussions}---}
In summary, we find a direct and continuous FCI-SF($\Gamma$) transition in hard-core boson Haldane model on the honeycomb lattice by simply tuning the bandwidth of the flat-band. The obtained critical exponents $\beta\approx0.35(5)$ and $\nu\approx0.62(12)$ are not only compatible with those of the 3D XY universality class, but also have room for more exotic ones beyond Landau-Ginzburg paradigm.  
Although the single-particle energy at $\Gamma$ and $M$ both changes linearly and monotonically with $t''$, the FCI-SF($M$) transition by tuning $t''$ as well turns out to be first-order.
We think the difference of the two transitions is that the bosons in SF($M$) state from our results spontaneously condense at one single $M$ point and the $C_3$ rotational symmetry is broken as well. In contrast, for the FCI-SF($\Gamma$) transition, there is only $U(1)$ symmetry breaking. 
However, it is still an interesting open question to study the phase transition from an isotropic FCI to a fragmented SF($M$) ~\cite{Ho2000fragment,Mueller2006fragment} where the bosons equally condense at all $M$ points or the possible phase transition from anisotropic/nematic FCI to the rotational-symmetry-broken SF($M$).

Furthermore, it is also intriguing to study the $V_1(V_2)-t''$ phase diagram with even larger neighboring interactions since there will be phase transitions from FCI to translational symmetry breaking states at the flat-band limit~\cite{DNSheng2011boson, Luo2020BosonFCI, Lu2024vestigial}. Considering more dispersive Chern bands when tuning $t''$, the phase diagram would become more complex and might include possible supersolid states. For example, it is interesting to study whether the roton modes in FCI would survive in the continuously obtained SF($\Gamma$) state, and whether the supersolid could be (quasi)adiabatically realized by softening the possible roton mode in SF($\Gamma$), realizing a sequence of continuous FCI-SF-supersolid transitions, which we leave for future investigations. We notice the softening of roton modes in SF is a proposed scheme for the preparation of supersolids being intensively studied, such as in dipolar gas systems~\cite{Chomaz2018roton, Tanzi2019supersolid, Fabien2019supersolid,Chomaz2019supersolid, zhang2019supersolid, Alana2023supersolid}.

The continuous FCI-SF($\Gamma$) transition, potentially of ``beyond Landau'' type~\cite{Barkeshli2012CFL_FL, Barkeshli2014FQH-SF, Barkeshli2015FCI-SF,Song2023deconfined, songPhase2024,senthil2004DQCP,senthilQuantum2004}, would have many other exotic yet detectable features as well, which are meaningful for future investigations.
For example, we notice that the charge-density-wave (CDW) fluctuations and emergent SO(3) symmetry are predicted at the critical point of the continuous transitions between composite Fermi liquid and Fermi liquid and between chiral spin liquid and topological superconductors~\cite{Song2023deconfined, songPhase2024}.

Experimentally, this work further suggests the feasibility and provides new perspectives of the (quasi)adiabatic preparation of FCI states from SF states in ultracold atom systems~\cite{Barkeshli2015FCI-SF}. And we believe the realization of zero-field bosonic FCI in a Haldane-model Hamiltonian is feasible since the up to NNNN hoppings and non-trivial single-particle topology can be realized in optical lattices by periodic modulation on the phases of the lattice beams and the on-site as well as NN interactions can be realized by conventional s-wave Feshbach resonance~\cite{Chin2010Feshbach,Marin2015periodic, Miao2022boson}. Our $V_1(V_2)-t''$ phase diagram should suggest versatile routes of such continuous preparations.
With the great achievements in realizing topological models with different geometries
using ultracold systems and circuit
quantum electrodynamics techniques~\cite{ Cooper2019ultracold, Tai2017Hofstadter,Owens2018QED, Owens2022QED, palm2024growing}, it might be worthwhile to
numerically check whether at least the continuous transition between FCI and SF (breaking only $U(1)$ symmetry) states could apply to other topological models with different geometries by similarly tuning the band dispersion.

\begin{acknowledgments}
{\it Acknowledgments}\,---\, We thank Kai Sun, Meng Cheng, Chong Wang, Senthil Todadri for helpful discussions. ZYM thanks Zhendong Zhang for discussions on ultracold atoms. HYL thanks Xue-Yang Song for helpful discussions. 
We use QSpace package for the finite DMRG simulations~\cite{AW2012_QSpace} and TeNPy package for the infinite DMRG simulations~\cite{tenpy2018}. HYL, BBC and ZYM acknowledge the support from the Research Grants Council (RGC) of Hong Kong (Project Nos. 17301721, AoE/P-701/20, 17309822, HKU C7037-22GF, 17302223, 17301924), the ANR/RGC Joint Research Scheme sponsored by RGC of Hong Kong and French National Research Agency (Project No. A\_HKU703/22). We thank HPC2021 system under the Information Technology Services and the Blackbody HPC system at the Department of Physics, University of Hong Kong, as well as the Beijng PARATERA Tech CO.,Ltd. (URL: https://cloud.paratera.com) for providing HPC resources that have contributed to the research results reported within this paper. 
H.Q. Wu acknowledge the support from GuangDong Basic and Applied Basic Research Foundation (No. 2023B1515120013) and Youth S$\&$T Talent Support Programme of Guangdong Provincial Association for Science and Technology (GDSTA) (No. SKXRC202404). The ED calculations reported were performed on resources provided by the Guangdong Provincial Key Laboratory of Magnetoelectric Physics and Devices (No. 2022B1212010008).
\end{acknowledgments}

\bibliographystyle{apsrev4-2}

\newpage\clearpage
\renewcommand{\theequation}{S\arabic{equation}} \renewcommand{\thefigure}{S%
	\arabic{figure}} \setcounter{equation}{0} \setcounter{figure}{0}

\begin{widetext}
	
\section{Supplementary Information for \\[0.5em]
Continuous Transition between Bosonic Fractional Chern Insulator and Superfluid}
In the Supplementary Information, we further show the absence of CDW orders in SF states, the entanglement entropy scaling in the SF($\Gamma$) state, the topologically trivial behavior of the SF states, the correlation length of the continuous FCI-SF($\Gamma$) transition, the supplementary ED results, and the first-order FCI-SF($K$) transition with attractive interactions.

\subsection{Section I: Absence of CDW order in the SF states}

The $\nu=1/2$ FCI state in this model has been well studied and there is no CDW order in the parameter-rigme that we have considered in this work~\cite{DNSheng2011boson, Luo2020BosonFCI}. In this section, we show that there is no CDW order in the SF states either.

\begin{figure}[htp!]
	\centering		
	\includegraphics[width=0.89\textwidth]{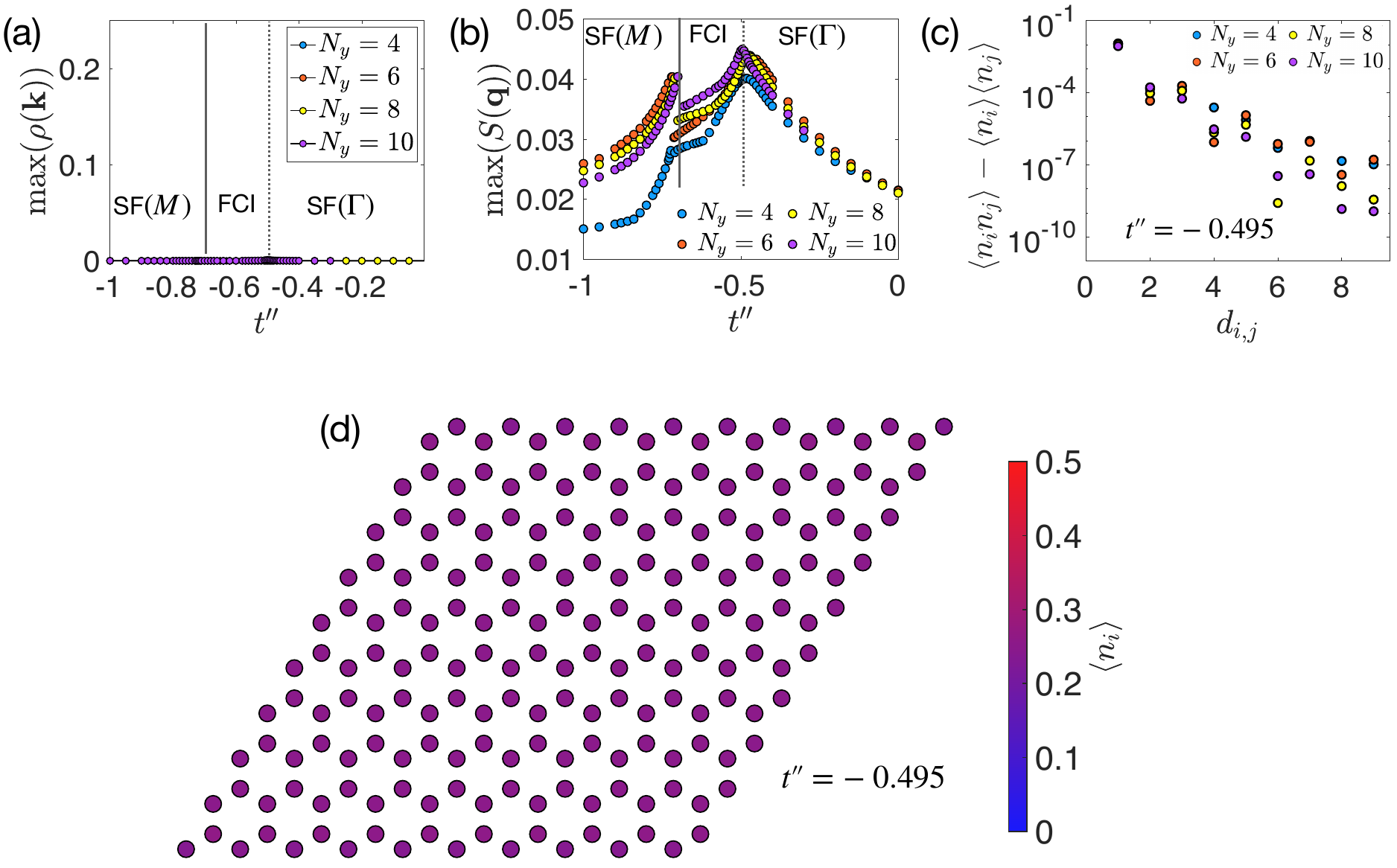}
	\caption{ \textbf{Absence of CDW order with $V_1=V_2=0$.} (a) The maximum of $\rho(\mathbf{k})$ as a function $t''$. (b) The maximum of $S(\mathbf{q})$ as a function $t''$. (c) The density-density correlations at $t''=-0.495$ as a function of distance $d_{i,j}$ between two sites, showing the exponential-decay behavior. (d) Uniform real-space boson distribution of $10\times10\times2$ sites in the bulk of a $N_y=10$ cylinder at $t''=-0.495$. }
	\label{fig_figS1}
\end{figure}	

We define $\rho(\mathbf{k})=\sum_{i,j}e^{-i\mathbf{k}\mathbf{r_i}}(\langle n_{i,\alpha}\rangle-\bar{n})/N'$ to detect the real-space distribution of the bosons, where $\alpha$ refers to the sublattices and we take B sublattice for example. Still, we do the Fourier transformation in a bulk region of $N'=N_y\times N_y$ sites. If there is any spontaneously translational-symmetry-breaking, there will be Bragg peaks in $\rho(\mathbf{k})$ in the order of $\bar{n}=0.25$ (half-filling the lower Chern band). As shown in Fig.\ref{fig_figS1}(a), the maximum of $\rho(\mathbf{k})$ is almost 0 across the $V=0$ phase diagram when tuning $t''$. 
Therefore, the translational symmetry is not spontaneously broken in the SF states either.
To further rule out the possibility of CDW orders, we define the static structure factor $S(\mathbf{q})=\frac{1}{N'}\sum_{i,j}e^{-i\mathbf{q}{r_{i,j}}}(\langle n_{i,\alpha}n_{j,\alpha}\rangle-\langle n_{i,\alpha}\rangle\langle n_{j,\alpha}\rangle)$ (the way we do Fourier transformation is the same as that for $n(\mathbf{k})$) and show its maximum value as a function of $t''$ in Fig.\ref{fig_figS1}(b). Such small values of $S(\mathbf{q})$ further verify the absence of long-range order or density-density correlations across the phase diagram. 
Furthermore, the density fluctuations are even decreasing with system size in the SF states (except the finite-size effect for $N_y=4$).
Besides this, these structure factors still provide some other information about the density fluctuations. For example, its evolution is continuous across the FCI-SF($\Gamma$) transition while it is more and more discontinuous across the FCI-SF($M$) transition when system size increases.

We notice that the density fluctuations are relatively stronger around the FCI-SF($\Gamma$) transition, so we provide further information of the density-density correlations. We plot the absolute value of $\langle n_in_j\rangle$ as a function of the distance between the two sites at $t''=-0.495$ in Fig.\ref{fig_figS1} (c). The density-density correlations show clear exponential decay for all system sizes. 
We also show the uniform real-space boson distribution in the bulk of $10\times10\times2$ lattice sites at $t''=-0.495$ to further demonstrate the translational invariance.


\begin{figure}[htp!]
	\centering		
	\includegraphics[width=0.89\textwidth]{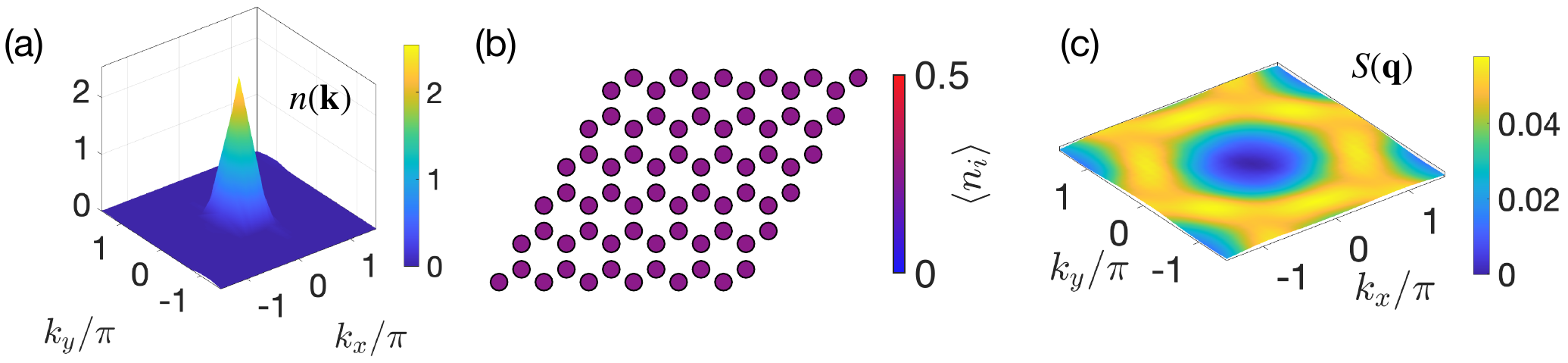}
	\caption{ \textbf{Absence of CDW order with strong neighboring interactions.} We take $V_1=4.8$, $V_2=1.8$, and $t''=-0.08$ as an example of the SF($\Gamma$) state, which refers to the strongest neighboring interactions considered in this work. 
	(a) The momentum-space boson occupation in the bulk of a $N_y=6$ cylinder (the Fourier transformation is conducted among the central $6\times6=36$ sites to obtain $n(\mathbf{k})$). (b) The uniform real-space boson distribution. (c) The static structure factor without any Bragg peaks, showing the absence of CDW order or strong density fluctuations.}
	\label{fig_figS2}
\end{figure}	

We have introduced in the main text that the continuous FCI-SF($\Gamma$) transition extends to parameter regions with strong neighboring interactions, and there is no CDW order in the SF($\Gamma$) state within the considered parameter regimes. We take the strongest interactions ($V_1=4.8$ and $V_2=1.8$) that are considered in this work as an example for demonstration. To rule out the possible CDW in the SF($\Gamma$) state with strong interactions, we show the results at $t''=-0.08$ and the boson condensation is shown in Fig.\ref{fig_figS2}(a).
The uniform boson distribution in the real-space, as shown in Fig.\ref{fig_figS2}(b) supports that the translational symmetry is preserved.
Besides, we show the static structure factor in Fig.\ref{fig_figS2}(c), and there are no Bragg peaks or strong density fluctuations.

The interplay of CDW orders between these states might be interesting for future investigations by further considering larger neighbor interactions.

\subsection{Section II: Entanglement entropy in the SF($\Gamma$) state}
In the main text, we have shown that the entanglement entropy at the critical point of the FCI-SF($\Gamma$) transition satisfies the area law without the logarithm dependence of $x$, which supports that our DMRG results well capture the (2+1) D nature of this continuous transition. 
Here, we provide the supplementary data of entanglement entropy in the SF($\Gamma$) state and take $V_1=V_2=t''=0$ as an example.

\begin{figure}[htp!]
	\centering		
	\includegraphics[width=0.6\textwidth]{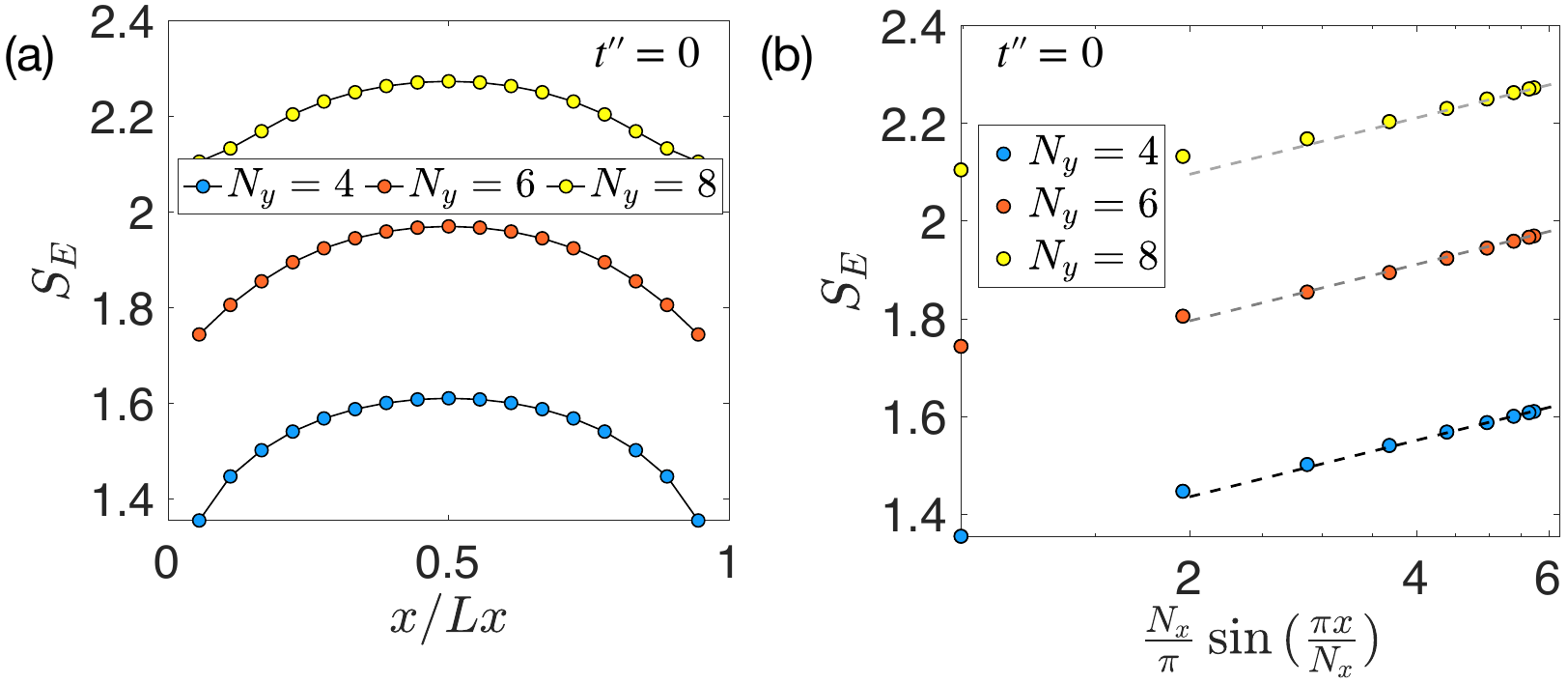}
	\caption{ \textbf{Entanglement entropy of SF($\Gamma$) at $V_1=V_2=t''=0$.} (a) The entanglement entropy for $N_y=4,6,8$ cylinders, which clearly show the logarithm dependence of $x$. (b) The entanglement entropy versus conformal distance (defined as $\frac{N_x}{\pi}\sin(\frac{\pi x}{N_x})$). The slope of the dashed lines with the respect to the conformal distance is $1/6$, showing the central charge $c=1$.}
	\label{fig_figS3}
\end{figure}	

As shown in Fig.\ref{fig_figS3}, apart from the $N_y$-dependence, the logarithm dependence of $x$ for all $N_y$ is clear and the fitted central charge is $c=1$ from $S_\mathrm{E}(x)\sim\frac{c}{6}\ln[\frac{N_x}{\pi}\sin(\frac{\pi x}{N_x})]$ for every $N_y$.

\clearpage

\subsection{Section III: Topologically trivial SF states}

\begin{figure}[htp!]
	\centering		
	\includegraphics[width=0.8\textwidth]{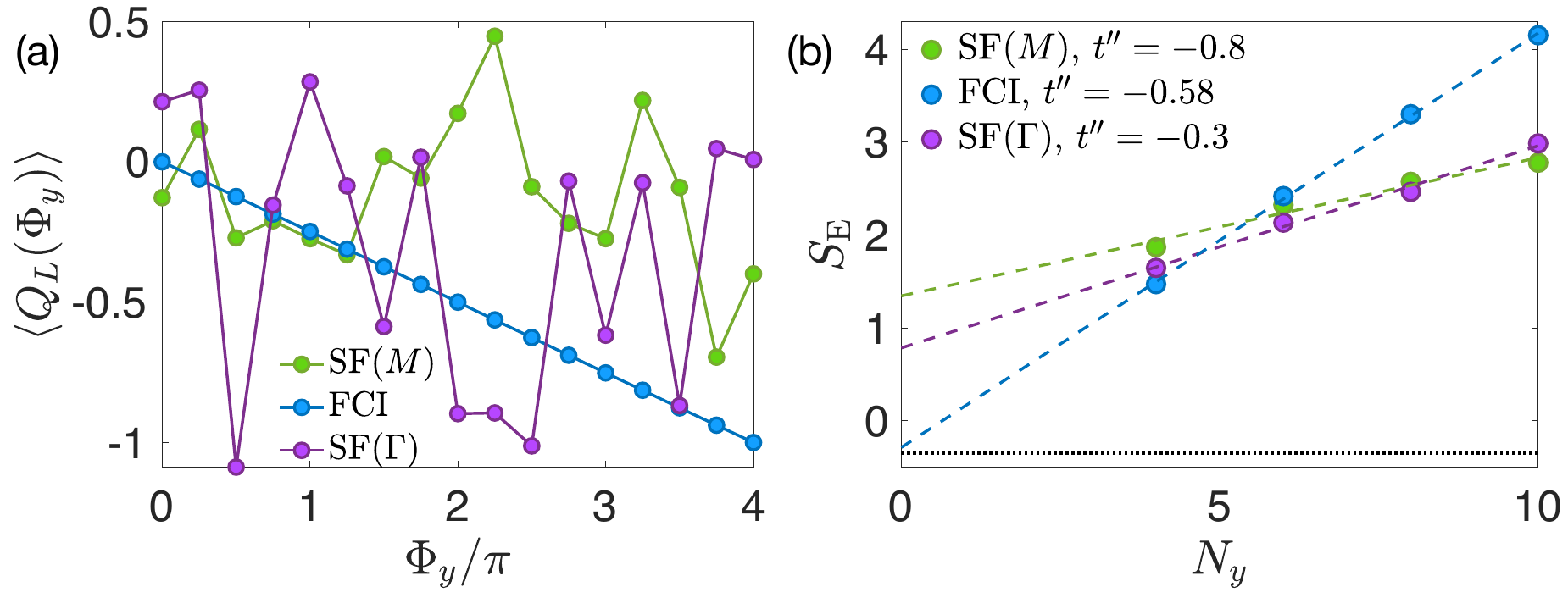}
	\caption{ \textbf{Charge pumping and topological entanglement entropy.} (a) Charge pumpings of the three states with no neighboring interactions ($V_1=V_2=0$) aftering adiabatically inserting $4\pi$ fluxes show the fractionally quantized Hall conductivity $\sigma_{xy}=0.5e^2/h$ for the FCI ($t''=-0.58$), and unquantized values for SF($M$) with $t''=-0.8$ and SF($\Gamma$) with $t''=-0.3$. The accumulated charges on the left edge ($Q_L$) of the cylinders are counted.
	(b) The extrapolations of bipartite entanglement entropy versus $N_y$ with the same $N_x$ show that only the FCI has the nontrivial topological contribution with the extrapolated value around $-0.29$, which is quite close to the black dotted line with the value $-\log(\sqrt{2})$.
 }
	\label{fig_figS_topology}
\end{figure}	

As introduced in the main text, we are studying quantum phase transitions between a topologically ordered FCI and topologically trivial SF states, and we provide more numerical evidence to support that the SF states are indeed topologically trivial.

The gapless energy spectrum from ED simulations are shown in Fig.\ref{fig_twist}, and we focus on the DMRG results in this section. 
First, we measure the Hall conductivities in all the three states with $V_1=V_2=0$ by adiabatically inserting fluxes in an infinite cylinder with $N_y=6$~\cite{Grushin2015_FCI_iDMRG}.
As shown in Fig.\ref{fig_figS_topology}(a), aftering adiabatically inserting $4\pi$ fluxes, 1 boson is pumped from the left to the right edge of the cylinder, indicating the fractionally quantized Hall conductivity $\sigma_{xy}=0.5e^2/h$. 
Although the time-reversal symmetry is explicitly broken, the topologically trivial SF states show unquantized $\sigma_{xy}$, which is consistent with previous ED results of topologically trivial SF states with broken time-reversal symmetry by calculating the Berry curvature~\cite{Barkeshli2015FCI-SF}.

Furthermore, we extrapolate the bipartite entanglement entropy from cylinders with the same $N_x$ and different $N_y$.  By doing so, for states with the same topological order as the $\nu=1/2$ bosonic Laughlin state, the entanglement entropy will scale as $S_\mathrm{E}=aN_y-\gamma$, where $\gamma=\log(\sqrt{2})$ is known as the topological entanglement entropy~\cite{Kitaev2006_topological_entanglement_entropy, Grushin2015_FCI_iDMRG}. As shown in Fig.\ref{fig_figS_topology}(b), the extrapolated value of the FCI state is around $-0.29$ and quite close to $-\gamma$, which verifies its topological order. However, although the time-reversal symmtry is explicitly broken and the Chern band is half filled, the superfluid states obviously do not have such topological contributions, which further rules out the possibility of coexisting topological order in the SF states.

\clearpage
\subsection{Section IV: Correlation length of the continuous FCI-SF($\Gamma$) transition}

\begin{figure}[htp!]
	\centering		
	\includegraphics[width=0.5\textwidth]{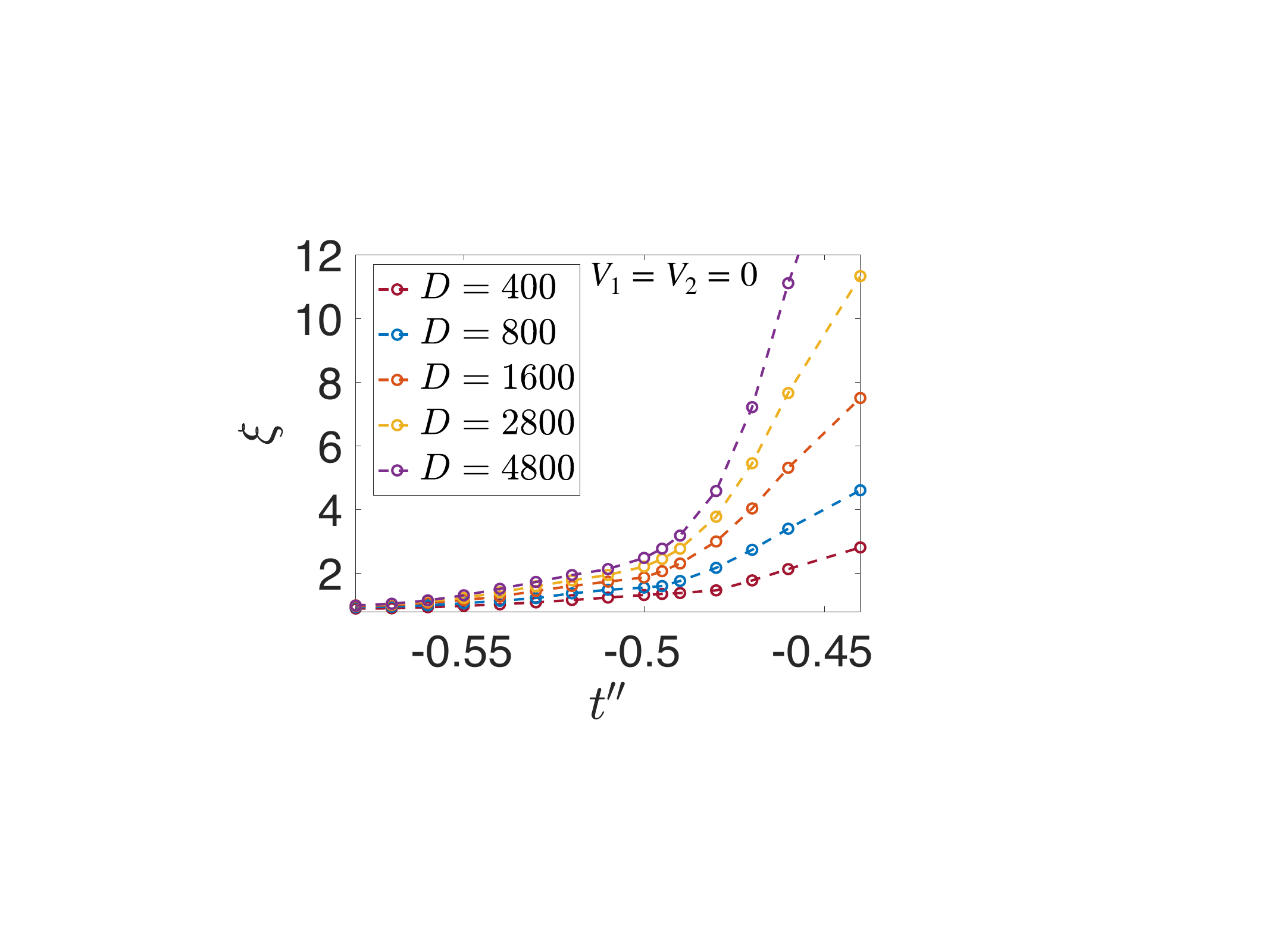}
	\caption{ \textbf{Correlation length of the FCI-SF($\Gamma$) transition.} Results are obtained from iDMRG similuations of $N_y=6$ systems across this continuous transition with $V_1=V_2=0$. The bond dimensions of the simulations are up to $D=4800$.}
	\label{fig_figS4}
\end{figure}	

To demonstrate the behavior of the correlation length $\xi$ of the FCI-SF($\Gamma$) transition, we implement infinite DMRG simulations (iDMRG) and obtain $\xi$ by diagonalizing the transfer matrix~\cite{tenpy2018}, instead of fitting the two-point correlation functions from the finite DMRG simulations. 
We take an unit cell of $N_y=6$ and $N_x=1$ (thus 12-leg infinite cylinder) for the simulations, and the results are shown in Fig.\ref{fig_figS4}. In the insulating FCI state, the correlation length is rather small and does not increase with bond dimension (converged). Approching the critical point and in the SF($\Gamma$) state, the correlation length keeps increasing with bond dimension, in agreement with the critical nature. At finite bond dimension, since the formation of off-diagonal long-range order, the correlation length in SF($\Gamma$) state increases faster than that at the critical point.
We also note that, for a fixed and finite bond dimension, the evolution (increase) of $\xi$ seems continuous across this transition from FCI to SF($\Gamma$). 
And this behavior is different from those transitions between two insulators where the correlation length diverges only at the critical point.

\subsection{Section V: Supplmentary results from ED}
\begin{figure}[htp!]
	\centering		
	\includegraphics[width=0.65\textwidth]{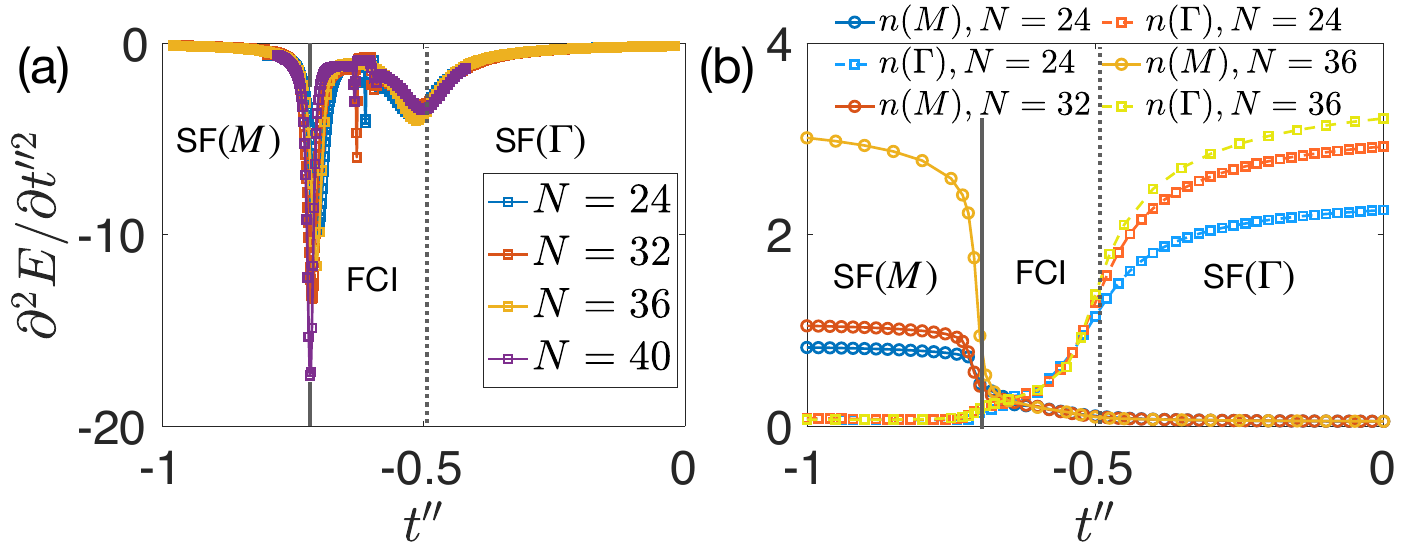}
	\caption{ \textbf{ED results of the phase transitions with $V_1=V_2=0$.} (a) The second derivative of per-site energy and (b) the momentum-space boson occupation number as functions of $t''$ for different tori, and the data points of ED results are taken very densely. 
	The solid and dashed lines refer to the two phase boundaries determined from the DMRG results. The singular values of $\partial^2E/\partial t''^2$ inside the FCI state are due to the cross of the 2-fold ground states themselves under the periodic boundary conditions and do not refer to any phase transition. In pabel (b), the $n($M$)$ of $N=36$ is much higher than those of other system sizes due to the different geometries of tori (the $N=36$ torus includes only one $M$ point in the BZ while the other tori include all $M$ points). }
	\label{fig_figS5}
\end{figure}

In this section, we show the ED results of the phase transitions with $V_1=V_2=0$ including the second derivative of per-site energy and the momentum-space boson occupation number in Fig.\ref{fig_figS5}. 
Overall, the ED results suffer from a severe finite-size effect, since it is hard to totally determine whether the FCI-SF(M) transition is first-order or not. This is in agreement with the DMRG results in Fig.\ref{fig_fig2} that the nature of first-order transitino is hard to distinguish from the results of $N_y=4$ cylinders. And the larger the system size, the first-order nature of the FCI-SF($M$) transition is more obvious.
But the ED results still show some difference of the two transitions from the behavior of $\partial^2E/\partial t''^2$ in Fig.\ref{fig_figS5}(a). Unlike those of the FCI-SF($\Gamma$) transition, although the values (second derivative of the energy per site) at the FCI-SF($M$) transition point are not discontinuous when the data points are taken very densely, they diverge too fast for such small system sizes, suggesting weakly first-order transitions.

\begin{figure}[htp!]
	\centering		
	\includegraphics[width=0.68\textwidth]{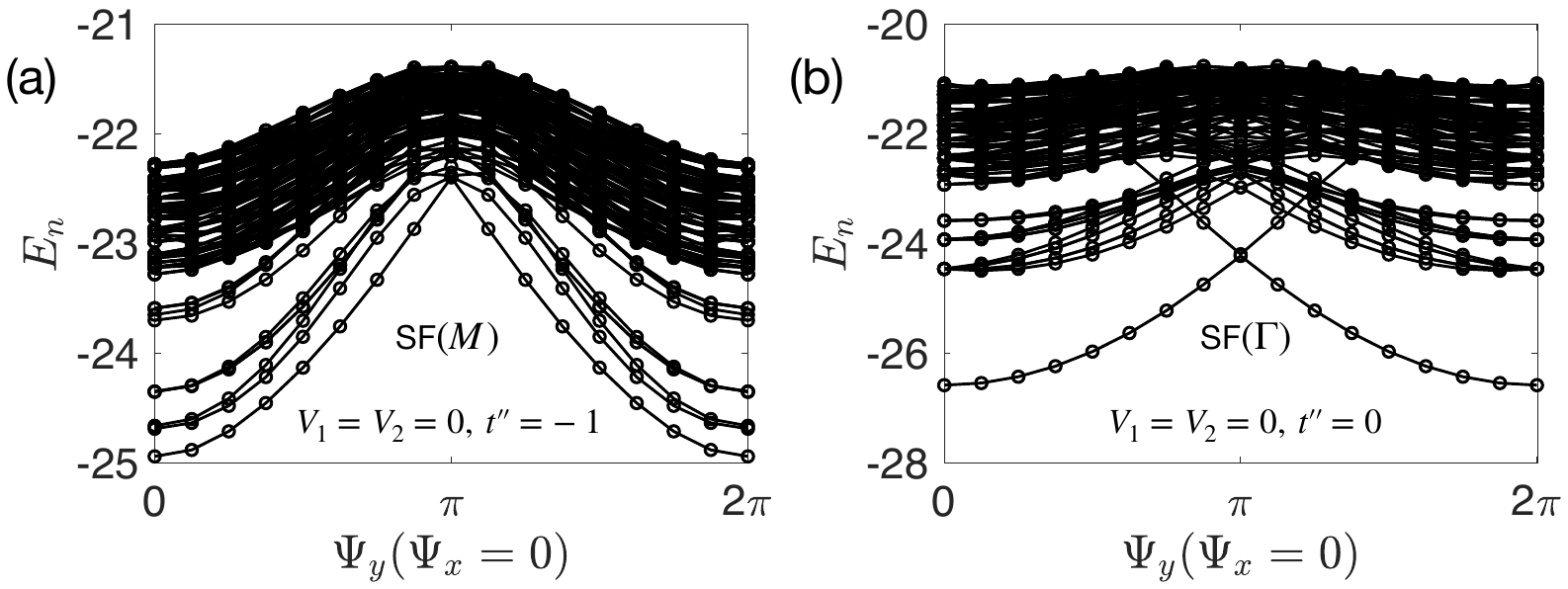}
	\caption{The gapless energy spectra of the two SF states under twisted boundary conditions from ED simulations of a $N=32$ torus.}
	\label{fig_twist}
\end{figure}

As the gapped ED spectrum of the FCI has been shown in Ref.\cite{DNSheng2011boson}, we show the spectra of the two SF states from ED simulations in Fig.\ref{fig_twist}, which clearly supports the gapless behavior of both SF states.

\subsection{Section VI: FCI-SF(K) transition}
	
\begin{figure}[htp!]
	\centering		
	\includegraphics[width=0.8\textwidth]{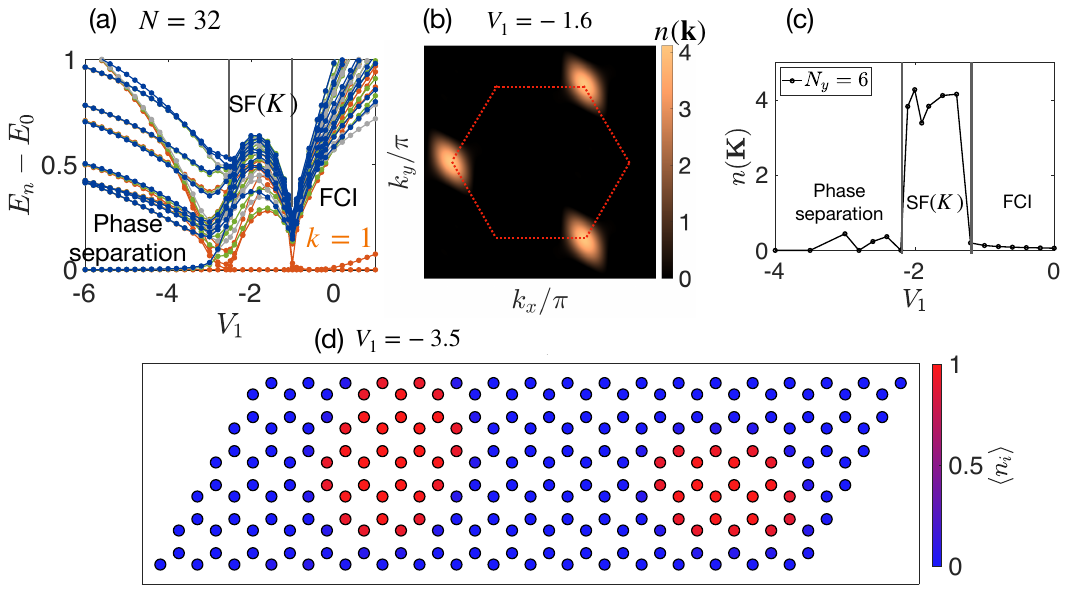}
	\caption{ We show the phase diagram with flat-band parameter ($t''=-0.58$) and attractive $V_1$. (a) Energy spectrum obtained by ED simulations of a $N=32$ torus, and the orange points are from the momentum sector $k=1$ ($\Gamma$ point in the BZ). (b) $n(\mathbf{k})$ obtained from DMRG simulations of a $N_y=6$ cylinder at $V_1=-1.6$. (c) Occupation $n(\mathbf{K})$ as a function of $V_1$ from DMRG simulations. (d) Real-space distributions of bosons in the phase separation with $V_1=-3.5$.}
	\label{fig_sf_K1}
\end{figure}	

In the main text, we focus on the continuous FCI-SF($\Gamma$) transition and the first-order FCI-SF($M$) transition by tuning the band dispersion. Here, we show that with the same $\nu=1/2$ filling of the lower flat band (fixed $t''=-0.58$), when gradually switching on the attractive NN interaction, there is another SF state with bosons condensed at $\mathbf{K}$ point in the Brillouin zone. The ED spectrum of a $N=32$ and change of $n(\mathbf{K})$ from DMRG simulations of a $6\times18\times2$ cylinder as functinos of $V_1$ are shown in Fig.\ref{fig_sf_K1}(a,c) respectively. We show the occupation of the SF($K$) phase in the momentum space with $V_1=-1.6$ as an example in Fig.\ref{fig_sf_K1}(b). The direct and first-order transition between FCI and SF($K$) is clearly shown from the ground-state level crossing and the jump of occupation $n(\mathbf{K})$. When the attractive interaction is further strengthend, there is a phase separation and we show the real-space occupation pattern with $V_1=-3.5$ in Fig.\ref{fig_sf_K1}(d) as an example.

\end{widetext}

\end{document}